\documentclass[epj]{svjour}
\usepackage{epsfig}
\usepackage{wrapfig}
\usepackage{amsmath}
\begin{document}
\newcommand{\myx}{\ensuremath{x} }
\newcommand{\qsq}{\ensuremath{Q^2} }

\def\Journal#1#2#3#4{{#1} {\bf #2} (#3) #4}
\def\NCA{\em Nuovo Cimento}
\def\NIM{\em Nucl. Instrum. Methods}
\def\NIMA{{\em Nucl. Inst. Meth.} {\bf A}}
\def\NPB{{\em Nucl. Phys.}   {\bf B}}
\def\PLB{{\em Phys. Lett.}   {\bf B}}
\def\PRL{{\em Phys. Rev. Lett.}}
\def\PRD{{\em Phys. Rev.}    {\bf D}}
\def\ZPC{{\em Z. Phys.}      {\bf C}}
\def\EJC{{\em Eur. Phys. J.} {\bf C}}
\def\CPC{{\em Comp. Phys. Commun.}}
\title{Studying Low-\boldmath{$x$} Dynamics using the Hadronic Final State in DIS at HERA}
\author{Roman P\"oschl 
\thanks{For the H1 and ZEUS Collaborations}
\thanks{Talk presented at the EPS03, Aachen, July 2003}}
\institute{DESY, Notkestr.~85, D-22603 Hamburg
}
\date{Received: {\today} / Revised version: {\today}}
\abstract{
This article describes different approaches to investigate the behavior
of parton evolution in the proton by exploiting various aspects of the hadronic
final state produced in Deep Inelastic Scattering Events at HERA.
} 
\maketitle
\section{Introduction}
\label{intro}
Measurements of the hadronic final state in deeply inelastic $ep$
scattering (DIS) provide precision tests of 
quantum chromodynamics (QCD). At HERA data are collected over a wide range of
the negative four-momentum-transfer $Q^2$, the Bjorken variable $x$ and the transverse
momenta $p_{T}$ of hadronic final state objects. 
\begin{wrapfigure}[27]{l}{3.5cm}
\epsfig{file=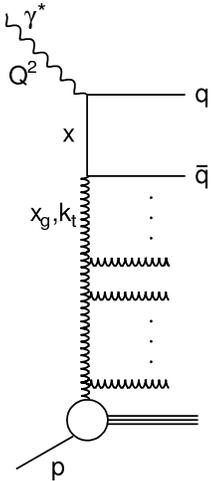, width=0.15\textwidth}
\caption{Diagram of a generic DIS process at low $x$. Here $k_t$ denotes the
transverse momenta of the exchanged gluons, $x_g$ is the fractional momentum
of the parton taking part in the hard interaction and $x$ is the Bjorken scaling variable}
\label{fig:1}
\end{wrapfigure}
Studies of the hadronic final state may be used to
get insight into the dynamics of the parton cascade exchanged in low-$x$
lepton proton interactions. 
Fig.~\ref{fig:1} shows a generic DIS process in which
a gluon from the proton undergoes a QCD cascade. 
The gluon interacts
with the virtual photon via a hard photon-gluon process which can be
calculated within perturbative QCD using an exact matrix element.
The cascade itself represents an approximation for an all order matrix
element calculation and  several prescriptions to describe the QCD
dynamics within the cascade have been proposed. 
The most familiar one is given by the so called DGLAP evolution equations~\cite{DGLAP}. In these
equations the large logarithms in $Q^2$ are resummed, neglecting
log$(1/x)$ terms. This 
practically corresponds to a strong ordering of the transverse momenta of the emitted partons,
$i.e.$ $k_{t,1}<<..<<k_{t,n}<<..<<Q^2$. DGLAP evolution is expected to
break down at sufficiently low values of $x$, when the ordering no
longer applies.

At very low values of $x$ it is
believed that the theoretically most appropriate description is given by the BFKL
evolution equations~\cite{FKL1}. These resum large logarithms of $1/x$ up 
to
all orders and impose no restriction on the ordering of the transverse
momenta within the parton cascade. Thus off-shell matrix
elements have to be used together with an unintegrated gluon distribution
function, $f(x,\tilde{\mu}_f^2,k_{t})$, which depends on the gluon transverse
momentum $k_t$ as well as $x$ and a hard scale $\tilde{\mu}_f$.  
A promising approach to parton evolution at both low and large values of $x$ is
given by the CCFM~\cite{CCFM} evolution equation, which, by means of
angular-ordered parton emission, is equivalent 
to the BFKL ansatz for $x
\rightarrow 0$, while reproducing the DGLAP equations at large $x$.

\section{Forward \boldmath{$\pi^0$}/Jet Cross Sections}\label{sec:1}
An extended parton ladder at low $x$ leads to high $k_t$ partonic
emission in the region close to the proton remnant ('forward' region) 
to which measurements of jets and leading particles, $e.g.$ $\pi^0$ are sensitive.
Production of a forward $\pi^0$ can be regarded as a refined
version of forward jet production. In order to enhance the sensitivity to low-$x$
effects special selection cuts have been applied such as confining the
ratio $p^{2}_{T,(\pi^0, Jet)}/Q^2$ to values between 0.5 and 2 inspired by
a proposal in~\cite{Mueller}.

\subsection{Forward \boldmath{$\pi^0$} Cross Sections} \label{subsec:11}
\begin{figure}
\centering
\epsfig{file=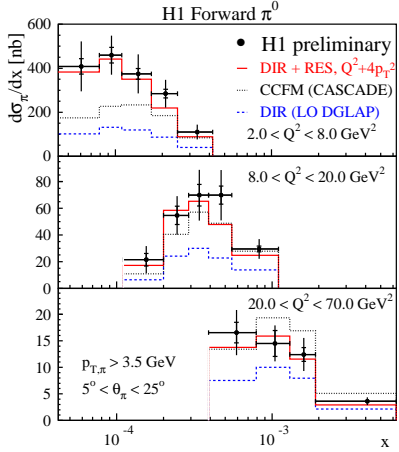, width=0.31\textwidth}
\caption{
Forward $\pi^0$ cross section as a function of Bjorken-$x$ in different regions
of $Q^2$ compared with predictions of DGLAP and CCFM based QCD Models.}
\label{fig:2}       
\end{figure}

Inclusive forward $\pi^0$ cross sections for transverse momenta
$p_{T,\pi^0} > 3.5$~GeV are shown in Fig.~\ref{fig:2} as a
function of $x$ for different regions of $Q^2$. The data
are compared with predictions of the Monte Carlo models RAPGAP~\cite{RAPGAP}
and CASCADE~\cite{CASCADE}. RAPGAP implements a QCD model based on Leading Order 
($O(\alpha_s)$, LO) parton showers with (`DIR +RES') and without (`DIR') resolved
photon structure. CASCADE is employed as an implementation of the CCFM evolution
equation introduced above. The prediction by RAPGAP with a pointlike
photon (DIR) is well below the data. A reasonable description of the
data is achieved by including contributions from resolved virtual photons
in the predictions and using a factorization scale of
$Q^2+4p_{T,\pi^{0}}$. Note, that resolved contributions can be
considered to mimic a lack of ordering in transverse momentum as required for
a DGLAP evolution scheme. CASCADE predictions based on the
unintegrated gluon density JS2001~\cite{CASCADE} on the other hand undershoot the data
for lower values of $Q^2$.

\subsection{Forward Jet Cross Sections} \label{subsec:12}

\begin{figure}
\centering
\epsfig{file=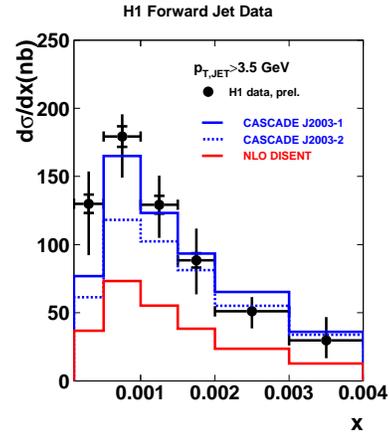, width=0.28\textwidth}
\caption{
Forward jet cross section as a function of Bjorken-$x$ compared with
NLO DGLAP QCD calculations and predictions by the CCFM Model CASCADE.}
\label{fig:3}       
\end{figure}

Results complementary to the ones discussed in Sec.~\ref{subsec:11} are obtained by
studying jets in the same region of phase space. Jets are
reconstructed with the longitudinally invariant $k_t$ cluster algorithm~\cite{INVKT}. 
Fig.~\ref{fig:3} shows the
forward jet cross section for transverse momenta $p_{T, Jet} > 3.5$~GeV 
as a function of $x$. The data are compared with NLO ($O(\alpha^{2}_s)$) QCD
calculations performed with the program DISENT~\cite{DISENT} and
predictions by CASCADE based on two recent sets of unintegrated gluon
distributions~\cite{JUNGDIS03}. While results of the NLO QCD calculations are significantly below
the data, the CASCADE  prediction based on the set labelled J2003-1 is 
in good agreement with the data. The difference between the CASCADE predictions
indicates the sensitivity of forward jet data to low-$x$ dynamics.

\section{Inclusive Jet Cross Sections}\label{sec:2}
\begin{figure}
\centering
\epsfig{file=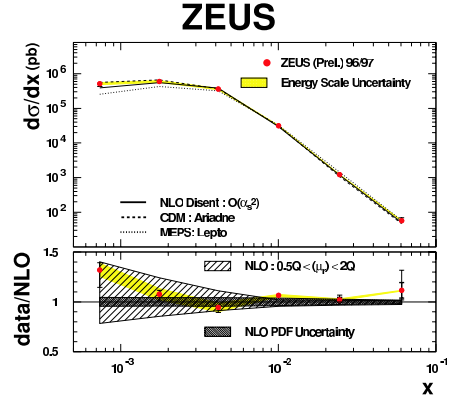, width=0.32\textwidth}
\caption{
Inclusive jet cross section as a function of Bjorken-$x$ compared with
NLO DGLAP QCD predictions and DGLAP based QCD Models.}
\label{fig:4}       
\end{figure}

In the following analysis of inclusive jet cross sections the restriction to
the forward region and the kinematic confinement introduced in Sec.~\ref{sec:1}
are removed, leading to a somewhat more general study of jet cross
sections. In~\cite{ZEUS} it is outlined that the comparison of the
measured jet cross sections with matrix element
calculations including contributions up to $O(\alpha_s)$ 
performed with DISENT lead to significant discrepancies between the data and
the theoretical predictions. If, however, hadronic activity in the
forward and the backward hemisphere is required and NLO ($O(\alpha^2_s)$)
predictions are employed a much better agreement
between the data and the theoretical prediction is obtained, which is
demonstrated in Fig.~\ref{fig:4}. The theoretical
error represented by the hatched band is due to missing higher order
contributions in the theoretical calculations. Predictions based on LO
DGLAP parton showers, here represented by LEPTO~\cite{LEPTO}, undershoot the data at
low-$x$ while a good description of the data is obtained by the CDM~\cite{CDM}
model as implemented in the event generator ARIADNE~\cite{ARIADNE}.

\section{Azimuthal Correlations between Jets}\label{sec:3}
Insight into low-$x$ dynamics can be gained from inclusive dijet data
by studying the behavior of events with a small azimuthal separation,
$\Delta\phi^{\ast}$, between the two hardest jets as measured in the hadronic
center-of-mass system~\cite{FORS,ASKE,SZCU}.  Partons entering the hard
scattering process with negligible transverse momentum, $k_t$, as assumed in the
DGLAP formalism, lead at leading order to a back-to-back configuration of the two outgoing
jets with $\Delta\phi^{\ast}=180^{\circ}$.  Azimuthal jet separations different 
from
$180^{\circ}$ occur due to higher order QCD effects.  However, in models which
predict a significant proportion of partons entering the hard process with large
$k_t$, the number of events with small $\Delta\phi^{\ast}$ should also
increase.

Here we present a measurement of the ratio
\begin{equation*}
S  =\frac{\int^{120^{\circ}}_{0}{N_{\rm dijet}(\Delta\phi^{*}, \myx,
\qsq){\rm d}\Delta\phi^{*}}} {\int^{180^{\circ}}_{0}{N_{\rm dijet}(\Delta\phi^{*
}, \myx,
\qsq){\rm d}\Delta\phi^{*}}}, 
\label{sasi}
\end{equation*}
of the number of events $N_{\rm dijet}$ with an azimuthal jet separation of
\mbox{$\Delta\phi^{\ast}< 120^{\circ}$} relative to all dijet events.
The observable was proposed in~\cite{SZCU} and is
considered to be directly sensitive to low-$x$ effects.

\begin{figure}
\centering
\epsfig{file=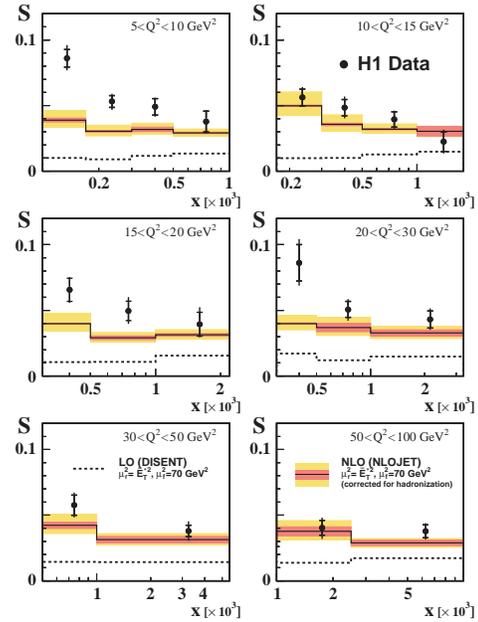, width=0.34\textwidth}
\caption{
The observable $S$ given as a function of Bjorken-$x$ and
$Q^2$ compared with LO and NLO DGLAP QCD predictions.
}
\label{fig:5}       
\end{figure}

Fig.~\ref{fig:5} presents the $S$-distribution as a function
of \myx for different $Q^2$. It is
of the order of~5\% and increases with decreasing $x$.  This rise of $S$
is most prominent in the lowest \qsq bin, where the smallest values of \myx are 
reached. 
The NLO dijet QCD calculations, resulting in an effective LO prediction
for $S$,
 predict $S$-values of only~$\sim$1\% and show no
rise towards low $x$. 
NLO 3-jet predictions by the program NLOJET~\cite{NLOJET} which lead
to an effective NLO prediction for $S$
give a good description of the data at large $Q^2$ and large $x$,
but still fail to describe the increase towards low $x$, particularly in the
lowest $Q^2$ range. According to the discussion given in~\cite{MYPAPER}, 
predictions based on CCFM evolution as implemented in CASCADE lead to an 
improved overall agreement with the data in particular at low $x$ and $Q^2$.

\end{document}